\documentclass[conference,letterpaper]{IEEEtran}

\usepackage[utf8]{inputenc} 
\usepackage[T1]{fontenc}
\usepackage{url}
\usepackage{ifthen}
\usepackage{cite}
\usepackage[cmex10]{amsmath} 

\usepackage{cite}
\usepackage{amssymb}
\usepackage{dsfont}
\usepackage{graphicx}
\usepackage{algorithm}
\usepackage{algorithmic}
\usepackage{diagbox}
\usepackage{mathtools}
\newcommand{\len}{\ell}

\newtheorem{definition}{Definition}
\newtheorem{lemma}{Lemma}
\newtheorem{theorem}{Theorem}
\newtheorem{remark}{Remark}
\newtheorem{example}{Example}

\newenvironment{examplecont}[1]
{
  \renewcommand{\theexample}{\ref{#1} (continued)}%
  \addtocounter{example}{-1}
  \begin{example}
}
{
  \end{example}
}

\DeclareMathOperator*{\argmin}{arg\,min}

\begin{document}
\title{A Rate-Distortion Framework for Summarization}

\author{
  \IEEEauthorblockN{Enes Arda and Aylin Yener}
  \IEEEauthorblockA{INSPIRE@OhioState Research Center\\ 
                    Dept. of Electrical and Computer Engineering\\
                    The Ohio State University \\
                   arda.2@osu.edu, yener@ece.osu.edu}
}

\maketitle

\begin{abstract}
    This paper introduces an information-theoretic framework for text summarization. We define the summarizer rate–distortion function and show that it provides a fundamental lower bound on summarizer performance. We describe an iterative procedure, similar to Blahut–Arimoto algorithm, for computing this function. To handle real-world text datasets, we also propose a practical method that can calculate the summarizer rate-distortion function with limited data. Finally, we empirically confirm our theoretical results by comparing the summarizer rate–distortion function with the performances of different summarizers used in practice.
\end{abstract}

\section{Introduction}

Text summarization stands out as an important task in natural language processing, where the goal is to produce a concise (low rate) version of the original text while preserving its main ideas (low distortion) \cite{summarization-gupta}. The recent success of Large Language Models (LLMs) across diverse tasks has expanded the realm of summarization as well, enabling the generation of remarkably high-quality summaries \cite{summeval-2021, bhandari-etal-2020-evaluating}. Despite these advances in practice, the theoretical limits of summarizers remain largely unexplored.

Information-theoretic tools have already proven helpful in summarization. For example, reference \cite{darrin-etal-2024-cosmic} shows that the classification error probability based on the summary is bounded above and below by functions of mutual information. The authors propose using mutual information as a summarizer evaluation metric, and demonstrate its strong correlation with human judgement. Similarly, in reference \cite{kim2022mutual}, mutual information serves as a unified metric for multimodal generative models. Reference \cite{padmakumar-he-2021-unsupervised} uses mutual information to select the most informative sentences for extractive summarization, and reference \cite{Poel2022MutualIA} leverages it as an objective in abstractive summarization. However, none of these studies investigate the trade-off between the rate and the expected distortion in summarization.

Rate-distortion theory describes the fundamental trade-off between the average length of a compressed representation and the average distortion incurred during reconstruction \cite{shannon1959coding, berger1971rate}. In this framework, \emph{distortion} is defined between the reconstruction and the original input, while \emph{length} is only defined for the compressed representation. As a result, only the size of the compressed representation matters, not its exact form. By contrast, in summarization, both length and distortion are defined with respect to the original text and the summary. Hence, both the content of the summary and its length are essential.

Recently, there has been growing interest in a specific kind of summarization task known as prompt compression \cite{jiang-etal-2023-llmlingua}, where the goal is to keep the input prompt to an LLM as short as possible while still producing an output that is similar to the output of the original, longer prompt. In reference \cite{nagle2024fundamental}, this problem is described as a distortion-rate function, formulated as a linear program, and an algorithm is proposed to solve it. However, the proposed algorithm is feasible only when the compression is done by pruning. While pruning may be a suitable approach for prompt compression, it does not apply to summarization in general.

In this work, we aim to develop an information theoretic framework for text summarization, which follows an analysis similar to the classical rate-distortion theory. First, we define the summarizer rate-distortion function for a given text source and a distortion metric that measures the dissimilarity between the text and a summary. Then we show that this function provides a fundamental lower bound on the summarizer's rate for a specified average distortion. Next, we describe an iterative algorithm to compute the summarizer rate-distortion function. Finally, recognizing the challenges of applying this algorithm to real-world datasets, we propose a practical method that can effectively compute the summarizer rate-distortion function with limited data availability. Our experiments demonstrate that the summarizer rate-distortion function computed with the proposed method serves as a plausible lower bound on popular summarizers used in practice. Overall, our work offers an information theoretic viewpoint for text summarization and marks a step towards understanding the fundamental limits of summarizer capabilities.

\section{Problem Setup}

Let \(\mathcal{A}\) be a finite alphabet, the elements of which are referred to as letters. We consider a text source that generates a string of letters \mbox{\(T \sim p_T\)}, where \(T\) takes values in \(\mathcal{T} \subset \mathcal{A}^*\), and \(p_T\) is the probability distribution of \(T\). \(\mathcal{A}^*\) denotes the set of all finite strings constructed from the elements of \(\mathcal{A}\). A sequence of strings \(T^n=(T_1,\dotsc,T_n)\) is generated i.i.d. according to the distribution \(p_T\). For any \(t \in \mathcal{T}\), \(\len(t)\) denotes the length of the string. We assume that \(\mathbb{E}[\len(T)^3]\) is finite. For a sequence of strings, we have \(\len(t^n)\coloneqq\sum_{i=1}^{n}\len(t_i)\). \(p_{S|T}\) describes a Markov kernel and \(p_S\) is the marginal distribution induced by \(p_{S|T}\). To simplify the notation, we use \(p(t) \coloneqq p_T(t)\),  \(p_{S|l} \coloneqq p_{S|\len(T)=l}\), \(L_i \coloneqq \len(T_i)\) whenever there is no ambiguity. \(\Pr(A)\) denotes the probability of event \(A\).

\begin{definition} \label{def:summarizer}
A summarizer \(p_{S^n \mid T^n}\) is defined as a random mapping from \(n\) input strings \(t^n \in \mathcal{T}^n\) to \(n\) output strings \(s^n \in \mathcal{S}^n\), such that \(\Pr\left(\len(S_i) \leq \len(T_i)\right)=1\) for every \(i=1,2,\dotsc, n\). Here, \(\mathcal{S} \subset \mathcal{A}^*\) but it is not necessarily equal to \(\mathcal{T}\). We say that the summarizer is operating at \emph{rate} \(R \in \mathbb{R}\) if \(R\) is the smallest real number satisfying
\begin{equation}
    \Pr\left(\mathbb{E}\left[\frac{\len(S^n)}{\len(T^n)} \,\middle|\, \len(T^n)\right] \leq R\right) = 1.
\end{equation}
\end{definition}

\begin{definition}
A text-summary distortion function is a mapping 
\begin{equation}
  d: \mathcal{T} \times \mathcal{S} \mapsto \mathbb{R}^+,
\end{equation}
where \(\mathbb{R}^+\) is the set of non-negative real numbers. \(d(t, s)\) measures the \emph{dissimilarity} between the original text \(t \in \mathcal{T}\) and the summary \(s \in \mathcal{S}\). For a sequence of strings, we define \(d(t^n,s^n)\coloneqq\frac{1}{n}\sum_{i=1}^{n}d(t_i, s_i)\). Furthermore, we say that a distortion function is \emph{normal} if for every \(t \in \mathcal{T}\), there exists \(\hat{s}_t \in \mathcal{S}\) such that \(d(t, \hat{s}_t)=0\) (see Definition~8.5 in \cite{yeung-info-th}).
\end{definition}

To illustrate these definitions, we now present a simple example of one-shot summarizers.

\begin{example} \label{ex:one-shot}
    Consider the binary alphabet \(\mathcal{A}=\{\mathtt{0}, \mathtt{1}\}\) and let \(p_T\) be a text source that assigns uniform probabilities to the strings in \(\mathcal{T}=\{\mathtt{0000}, \mathtt{0010}, \mathtt{0110}, \mathtt{0111}\}\). Let \(\mathcal{S}=\{\mathtt{0}, \mathtt{10}, \mathtt{110}, \mathtt{111}\}\) be the set of summaries and define the distortion function as follows:

    \begin{equation} \label{eq:dist}
    \begin{array}{c|cccc}
      \hbox{\diagbox[width=2.5em]{$s$}{$t$}} & \mathtt{0000} & \mathtt{0010} & \mathtt{0110} & \mathtt{0111} \\ \hline
      \mathtt{0}   & 0    & 1    & 2    & 3    \\
      \mathtt{10}  & 5    & 0    & 1    & 2    \\
      \mathtt{110} & 5    & 5    & 0    & 1    \\
      \mathtt{111} & 5    & 5    & 5    & 0   
    \end{array}
    \end{equation}
    Notice that this distortion function is normal as the diagonal entries in \eqref{eq:dist} are all zero. A \emph{good} summarizer aims to minimize the expected distortion at a given rate. In this setup, we will construct three one-shot summarizers \(p_{S|T}\), each achieving a different \((D, R)\) pair, where \(D=\mathbb{E}_{p_{S|T}}[d(T,S)]\) is the expected distortion, and \(R\) is as defined in Definition~\ref{def:summarizer}. Notice that because the text length is \(4\) with probability \(1\), in this setup, we have \(R=\mathbb{E}_{p_{S|T}}\left[\frac{\len(S)}{\len(T)}\right]\).

    Let us first construct \(p^{(1)}_{S|T}\), which operates at the lowest possible rate. Since the shortest summary length in \(\mathcal{S}\) is \(1\), the lowest possible rate is \(0.25\), which is achieved when the summary is \(\mathtt{0}\) with probability 1. For this summarizer \(p_{S|T}^{(1)}\), the expected distortion is
    \begin{equation}
        \mathbb{E}_{p_{S|T}^{(1)}}[d(T,S)]      
        =
        \tfrac{1}{4}\times 0
        \;+\;
        \tfrac{1}{4}\times 1
        \;+\;
        \tfrac{1}{4}\times 2
        \;+\;
        \tfrac{1}{4}\times 3
        =
        1.5.
    \end{equation}
    Hence, \(p_{S|T}^{(1)}\) achieves the pair \((1.5, 0.25)\).

    Because the distortion function is normal, it is also possible to construct a summarizer that achieves zero distortion. Specifically, for each text, summarizer \(p_{S|T}^{(2)}\) selects the summary corresponding to the diagonal entry in \eqref{eq:dist} with probability \(1\). In this case, the achieved rate is
    \begin{equation}
        \mathbb{E}_{p_{S|T}^{(2)}}\left[\frac{\len(S)}{\len(T)}\right]
        =
        \tfrac{1}{4}\times \tfrac{1}{4}
        \;+\;
        \tfrac{1}{4}\times \tfrac{2}{4}
        \;+\;
        \tfrac{1}{4}\times \tfrac{3}{4}
        \;+\;
        \tfrac{1}{4}\times \tfrac{3}{4}
        \approx
        0.56,
    \end{equation}
    which yields the pair \((0, 0.56)\).
    
    Finally, consider the summarizer that selects between \(\mathtt{0}\) and \(\mathtt{111}\). It is easy to see from \eqref{eq:dist} that the lowest distortion is achieved when the summary is \(\mathtt{111}\) for text \(\mathtt{0111}\) and \(\mathtt{0}\) otherwise with probability \(1\), which achieves the pair \((0.75, 0.37)\).
    
    We constructed three summarizers that achieve the distortion-rate pairs $(0, 0.56)$, $(0.75, 0.37)$, and $(1.5, 0.25)$. Next in Sections~\ref{sec:fund-limits} and~\ref{sec:calculating-srd}, we calculate the fundamental lower bound for any summarizer $p_{S^n|T^n}$ in this setup, and compare with these values.
    
\end{example}

\section{Fundamental Limits of Summarizers} \label{sec:fund-limits}

In this section, we first define the summarizer rate–distortion function and then show that it serves as a lower bound on summarizer performance.

\begin{definition}
For a given text source \(p_T\), distortion function \(d: \mathcal{T} \times \mathcal{S} \mapsto \mathbb{R}^+\) and distortion constraint \(D \in \mathbb{R}^+\), the summarizer rate-distortion function is defined as

    \begin{equation} \label{def:summarizer-rd}
    \begin{aligned}
    R_S(D)\coloneqq\min_{p_{S|T}} \quad & \frac{1}{\bar{L}}I\left(T;S \mid \len\left(T\right)\right)\\
    \textrm{s.t.} \quad & \mathbb{E}\left[d\left(T, S\right)\right] \leq D,
    \end{aligned}
    \end{equation}
    where \(\bar{L} \coloneqq \mathbb{E}\left[\len\left(T\right)\right]\), \(I(T;S|L) \coloneqq \sum_lp(l)I(T;S|L=l)\), \(I(T;S|L=l) \coloneqq D_{\text{KL}}(p_{T,S|L=l} \| p_{T|L=l}p_{S|L=l})\), which implies \(I(T;S|L)=H(T|L)-H(T|S,L)\) in the discrete case. Note that all logarithms and exponentials are in base \(|\mathcal{A}|\).
\end{definition}

\begin{lemma} \label{lemma:mi_convexity}
    For a fixed \(p_T\), \(I\left(T;S \mid \len\left(T\right)\right)\) is a convex function of \(p_{S|T}\).
\end{lemma}

\begin{IEEEproof}
    Let \(\mathcal{T}_l\coloneqq\len^{-1}(l)\) be the inverse image of \(l \in \mathbb{N}\).
    \begin{equation}
        I\left(T;S \mid \len\left(T\right)\right)= \sum_{l}p(l)I\left(T;S \mid \len\left(T\right)=l\right).
    \end{equation}
    Since \(p(l)\) only depends on \(p_T\), it is sufficient to show that \(I\left(T;S \mid \len\left(T\right)=l\right)\) is a convex function of \(p_{S|T}\) for every \(l \in \mathbb{N}\).

    Fix \(l \in \mathbb{N}\). Since \(p_{T|\len(T)=l}\) only depends on \(p_T\), it is also fixed. For a fixed \(p_{T|\len(T)=l}\), \(I\left(T;S \mid \len\left(T\right)=l\right)\) is a convex function of \(p_{S|T,\len(T)=l}\) (see Theorem~2.7.4 in \cite{elements_of_info_th}). Since \(p_{S|T=t,\len(T)=l}=p_{S|T=t}\) for every \(t \in \mathcal{T}_l\), it follows that \(I\left(T;S \mid \len\left(T\right)=l\right)\) is a convex function of \(p_{S|T}\).
\end{IEEEproof}

\begin{lemma} \label{lemma:rd_convexity}
    \(R_S(D)\) is non-increasing and convex in \(D\).
\end{lemma}
\begin{IEEEproof}
    The minimization in \eqref{def:summarizer-rd} is over a set that grows with \(D\). Hence, \(R_S(D)\) is non-increasing in \(D\).

    Let \(p^{(1)}_{S|T}\) and \(p^{(2)}_{S|T}\) be two conditional pmfs that achieve \((D_1, R_S(D_1))\) and \((D_2, R_S(D_2))\). Define 
    \begin{equation}
        p^{(\lambda)}_{S|T}\coloneqq\lambda p^{(1)}_{S|T}+(1-\lambda)p^{(2)}_{S|T}.
    \end{equation}
    Since expectation is a linear function of the distribution, \(p^{(\lambda)}_{S|T}\) achieves the distortion \(D_{\lambda}=\lambda D_1 + (1-\lambda) D_2\). Also, let \(I_1, I_2 \) and \(I_\lambda\) be the conditional mutual information achieved with \(p^{(1)}_{S|T}, p^{(2)}_{S|T}\) and \(p^{(\lambda)}_{S|T}\) respectively. Then we can write
    \begin{align}
        R_S(D_\lambda) &\leq \frac{1}{\bar{L}}I_\lambda \label{eq:min_ineq}\\
                           &\leq \frac{1}{\bar{L}}(\lambda I_1 + (1-\lambda)I_2) \label{eq:convex_ineq} \\
                           &= \lambda R_S(D_1) + (1-\lambda)R_S(D_2),
    \end{align}
    where \eqref{eq:min_ineq} follows from the fact that \(p^{(\lambda)}_{S|T}\) is a feasible point for the minimization in \eqref{def:summarizer-rd}, and \eqref{eq:convex_ineq} follows from Lemma~\ref{lemma:mi_convexity}.
\end{IEEEproof}

We now present the main theorem, which states that the summarizer rate-distortion function establishes a fundamental lower bound on summarizer rate.

\begin{theorem} \label{thm:sum_bound1}
    Let a summarizer \(p_{S^n|T^n}\) be such that \(\mathbb{E}[\len(S^n)|\len(T^n)] \leq R\len(T^n)\) with probability \(1\). If \(\mathbb{E}\left[d\left(T^n, S^n\right)\right] \leq D\), then \(R \geq R_S(D)\).
\end{theorem} 

\begin{IEEEproof}[Proof Sketch]
Let \(\epsilon > 0\) and \(G=\mathds{1}\{\len(S^n) \leq R\len(T^n) + \epsilon n\}\). We first bound the conditional probability of the atypical event \(G=0\) given \(L^n = l^n\) using Chebyshev's inequality:

\begin{equation}
    \Pr(G=0 | l^n) \leq \frac{1}{4\epsilon^2n^2}\sum_{i=1}^n{l_i^2} \label{eq:prob_bound}.
\end{equation}

Depending on \(G\), we can calculate the number of possible summaries as we know the alphabet and the upper bound on the length of the summary. This yields for the typical event

\begin{equation}
    H(S^n|G=1, l^n) \leq R \sum_{i=1}^nl_i + \epsilon n + 1 \label{eq:cond_ent1},
\end{equation}
and similarly for the atypical event
\begin{equation}
    H(S^n|G=0, l^n) \leq \sum_{i=1}^nl_i + 1 \label{eq:cond_ent2}.
\end{equation}

Using \eqref{eq:prob_bound}, \eqref{eq:cond_ent1}, and \eqref{eq:cond_ent2}, we obtain the following upper bound on the entropy of \(S^n\) conditioned on \(L^n=l^n\):
\begin{align}
    H(S^n|l^n) &\leq \log{2} + R\sum_{i=1}^nl_i + \epsilon n + 1 \notag \\
    &\quad + \left(\frac{1}{4\epsilon^2n^2}\sum_{i=1}^n{l_i^2}\right) \left(\sum_{i=1}^nl_i + 1\right).
\end{align}
Taking expectation over text lengths gives
\begin{equation}
    H(S^n|L^n) \leq n(\bar{L}R + \epsilon) + \mathcal{O}(1), 
\end{equation}
which in turn yields
\begin{equation}
    R \geq R_S(D) - \frac{\epsilon}{\bar{L}} - \mathcal{O}\left(\frac{1}{n}\right).
\end{equation}

Now suppose that \(R_S^{(n)}(D)\) is the lowest rate achieved summarizing \(T^n\). Observe that the subsequence \(\{R_S^{(k)}(D)\}_{k=n,2n,4n,\dotsc}\) must be non-increasing, since the summarizer for \(T^n\) can be applied to \(T^{2n}\) in two blocks to achieve the same rate and distortion. Therefore, \(R \geq R_S^{(n)}(D) \geq \liminf_{k\to\infty}R_S^{(k)}(D) \geq R_S(D) - \frac{\epsilon}{\bar{L}}\). Because \(\epsilon\) was arbitrary, we have \(R \geq R_S(D)\).
\end{IEEEproof}

\section{Computing \(R_S(D)\)} \label{sec:calculating-srd}
In this section, we show that the summarizer rate-distortion function defined in \eqref{def:summarizer-rd} can be computed with a procedure similar to the Blahut-Arimoto algorithm \cite{blahut, arimoto}.

\begin{definition}
    Let 
    \begin{equation}
        s_l^* \coloneqq \argmin_{s \in \mathcal{S}} \mathbb{E}\left[d\left(T,s\right)|\len(T)=l\right].
    \end{equation}
    Then the maximum distortion for length \(l\) and the maximum distortion are defined as
    \begin{equation}
        D_{\text{max},l}\coloneqq\mathbb{E}\left[d\left(T,s_l^*\right)|\len(T)=l\right],
    \end{equation}
    \begin{equation}
        D_{\text{max}}\coloneqq\sum_{l}p(l)D_{\text{max}, l}.
    \end{equation}
\end{definition}

\begin{lemma} \label{lemma:zero-after-dmax}
    \(R_S(D)=0\) for \(D \geq D_{\text{max}}\).
\end{lemma}
\begin{IEEEproof}
    Let \(p_{S|T}\) be such that for a given \(\len(T)\), \(S=s_{\len(T)}^*\) with probability \(1\). Then \(I(T;S|\len(T)=l)=0\), and consequently \(I(T;S|\len(T))=0\). For this summarizer, we also have
    \begin{align}
        \mathbb{E}\left[d\left(T,S\right)\right]&=\sum_l{p(l)\mathbb{E}\left[d\left(T,s_l^*\right)|\len(T)=l\right]} \\
        &= \sum_{l}{p(l)D_{\text{max},l}} \\
        &= D_{\text{max}}.
    \end{align}
    Since \(R_S(D)\) is non-increasing by Lemma~\ref{lemma:rd_convexity}, \(R_S(D) \leq R_S(D_{\text{max}}) =0\) for \(D \geq D_{\text{max}}\).
\end{IEEEproof}

\begin{remark}
    In light of Lemma~\ref{lemma:zero-after-dmax}, we are only concerned with calculating \(R_S(D)\) for \(0\leq D \leq D_{\text{max}}\).
\end{remark}

\begin{lemma} \label{lemma:strictly-decreasing}
    If \(R_S(0) > 0\) then \(R_S(D)\) is strictly decreasing for \(0 \leq D \leq D_{\text{max}}\).
\end{lemma}
\begin{IEEEproof}
    Assume \(R_S(0) > 0\) and let \(R_S(D')=0\) for some \(D'\) which is achieved by some \(p_{S|T}\). Then for this \(p_{S|T}\), \(R_S(D')=\frac{1}{\bar{L}}I(T;S|\len(T))=0\), which implies \(I(T;S|\len(T)=l)=0\) for every \(l\). Hence, \(S\) and \(T\) are conditionally independent given \(\len(T)\), and \(p(t,s|l)=p(t|l)p(s|l)\). Then we can write
    \begin{align}
        D' &\geq \mathbb{E}\left[d\left(T,S\right)\right] \\
           &= \sum_{l}{p(l)\sum_{t,s}p(t,s|l)d(t,s)} \\
           &= \sum_{l}{p(l)\sum_{s}p(s|l)\sum_{t}p(t|l)d(t,s)} \\
           &= \sum_{l}{p(l)\sum_{s}p(s|l)\mathbb{E}\left[d\left(T,s\right)\right|\len(T)=l]} \\
           &\geq \sum_{l}{p(l)D_{\text{max},l}} \\
           &= D_{\text{max}}.
    \end{align}
    Therefore, \(R_S(D)>0\) for \(0 \leq D < D_{\text{max}}\). The facts \(R_S(D)>0\), \(R_S(D_{\text{max}})=0\) (Lemma~\ref{lemma:zero-after-dmax}), and \(R_S(D)\) is non-increasing and convex (Lemma~\ref{lemma:rd_convexity}) together imply that \(R_S(D)\) must be strictly decreasing for \(0 \leq D \leq D_{\text{max}}\).
\end{IEEEproof}

\begin{remark}
    If \(R_S(0)=0\), then the summarizer rate-distortion function is constant at zero. Therefore, we will assume \(R_S(0)>0\) without loss of generality. 
\end{remark}

Since \(R_S(D)\) is strictly decreasing for \(0 \leq D \leq D_{\text{max}}\) and it is convex, we can apply the same reasoning as in the Blahut-Arimoto algorithm (see Chapter~9.2.2 of \cite{yeung-info-th}). For any \(\beta \leq 0\), let \((D_{\beta}, R_S(D_{\beta}))\) be a point on the \(R_S(D)\) curve where the slope of the tangent at that point is equal to \(\beta\). Then, for this point, we have

\begin{align}
    R_S(D_\beta) - \beta D_\beta &= \min_{p_{S|T}}{\frac{1}{\bar{L}}I(T;S|\len(T))-\beta \mathbb{E}\left[d\left(T,S\right)\right]}\\
    \begin{split}
        &= \min_{p_{S|T}}\sum_{l}{p(l)}\bigl(\frac{1}{\bar{L}}I(T;S|\len(T)=l) \\ &\quad\qquad-\beta \mathbb{E}\left[d\left(T,S\right)|\len(T)=l\right]\bigr).
    \end{split} \label{eq:ba-final}
\end{align}
Notice that for every \(l\), the expression 
\begin{equation}
\frac{1}{\bar{L}}I(T;S|\len(T)=l)-\beta \mathbb{E}\left[d\left(T,S\right)|\len(T)=l\right]    
\end{equation}
in \eqref{eq:ba-final} depends on a different part of \(p_{S|T}\), namely \(\{p(\cdot|t)\}_{t \in \mathcal{T}_l}\) where \(\mathcal{T}_l=\{t | \len(t)=l\}\). Since \(\{\mathcal{T}_l\}_l\) form a partition for \(\mathcal{T}\), \(\{\{p(\cdot|t)\}_{t \in \mathcal{T}_l}\}_l\) form a partition for \(p_{S|T}\), and therefore we can write the optimization in \eqref{eq:ba-final} as
\begin{equation} \label{eq:ba-seperate-final}
    \begin{split}
    \sum_{l}{p(l)}\min_{\{p(\cdot|t)\}_{t \in \mathcal{T}_l}}\bigl(&\frac{1}{\bar{L}}I(T;S|\len(T)=l) \\ &-\beta \mathbb{E}\left[d\left(T,S\right)|\len(T)=l\right]\bigr).
    \end{split}
\end{equation} 

Because each minimization in \eqref{eq:ba-seperate-final} is of the same form as the objective Blahut-Arimoto algorithm minimizes, we can apply it for every \(l\) to calculate the summarizer rate-distortion curve, as described in Algorithm~\ref{alg:ba}.

\begin{algorithm}[ht]
\caption{Blahut-Arimoto for Summarizer Rate-Distortion}
\label{alg:ba}
\begin{algorithmic}[1]
\REQUIRE Source pmf $p_T$, distortion function $d(\cdot,\cdot)$, $\beta < 0$
\FOR{every $l$}
\STATE Initialize $\{q_l(s|t)\}_{t \in \mathcal{T}_l, s}$ arbitrarily
\REPEAT
  \STATE $r_l(s) \gets \sum_{t \in \mathcal{T}_l} p(t|l)\,q_l(s|t)$
  \STATE $q_l(s|t) \gets 
    \frac{r_l(s) \exp\left(\beta\,d\left(t,s\right)\right)}
         {\sum_{s} r_l(s) \exp\left(\beta\,d\left(t,s\right)\right)}$
\UNTIL{convergence}
\STATE $I_l \gets \sum_{t\in\mathcal{T}_l,s} p(t|l)\,q_l(s|t) \log{\frac{q_l(s|t)}{r_l(s)}}$
\STATE $D_l \gets \sum_{t\in\mathcal{T}_l,s} p(t|l)\,q_l(s|t) d(t,s)$
\ENDFOR
\STATE $R_S \gets \frac{1}{\bar{L}}\sum_{l} p(l)I_l$
\STATE $D \gets \sum_{l} p(l)D_l$
\end{algorithmic}
\end{algorithm}

\begin{figure}[htbp]
    \centering
    \includegraphics[width=\linewidth]{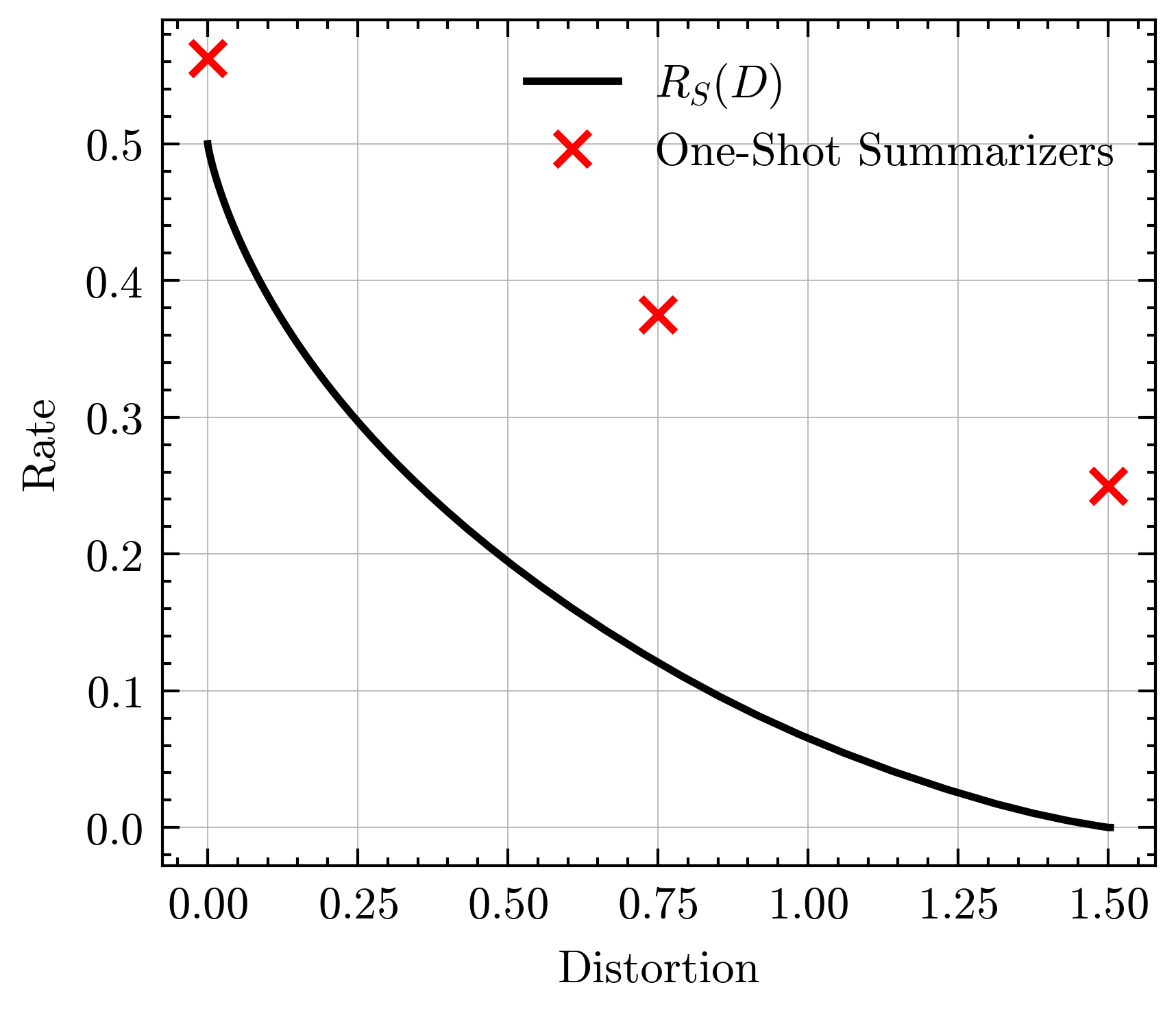}
    \caption{The summarizer rate-distortion curve, computed via Algorithm~\ref{alg:ba} for the setup of Example~\ref{ex:one-shot}, and the performance of one-shot summarizers.}
    \label{fig:example-rd}
\end{figure}

\begin{examplecont}{ex:one-shot}
    In this setup, the length of each generated text is always \(4\), and it can be seen from \eqref{eq:dist} that \(s_4^*=\mathtt{0}\). Consequently,
    \begin{equation}
      D_{\text{max}}
      =
      \tfrac{1}{4}\times 0
      \;+\;
      \tfrac{1}{4}\times 1
      \;+\;
      \tfrac{1}{4}\times 2
      \;+\;
      \tfrac{1}{4}\times 3
      =
      1.5.
    \end{equation}
    The function \(R_S(D)\) computed over \(0 \le D \le 1.5\) using Algorithm~\ref{alg:ba} is shown in Figure~\ref{fig:example-rd}. From the figure, we observe that \(R_S(1.5)=0\), that \(R_S(D)\) is strictly decreasing for \(0 \le D \le 1.5\), and that it serves as a lower bound on the rates of one-shot summarizers, all of which align with our theoretical findings.
\end{examplecont}

\section{Approximating \(R_S(D)\) From A Dataset}
In this section, we present a method for approximating the summarizer rate–distortion function on a real-world text dataset. We begin by identifying several challenges and then we describe how to address them.

\paragraph*{Estimating \(p_T\)}
A commonly used dataset for English summarization is CNN/DailyMail \cite{cnn_dailymail_dataset2015,cnn_dailymail_dataset2017}, which contains roughly 300k news articles. Different tokenization procedures yield different alphabets, but under the tokenization of \cite{cnn_dailymail_dataset2015}, the alphabet size is about \(|\mathcal{A}| \approx 100\text{k}\), and the average text length is \(\bar{L}\approx800\). Even if we consider a string of length \(10\), the number of possible texts is \(|\mathcal{A}|^{10}\approx10^{50}\). Since we have only 300k samples, it is infeasible to empirically estimate the source distribution over such a large space without extrapolation. The problem is compounded when estimating conditional distributions \(p_{T|\len(T)}\), as the data for any fixed text length is even more limited.

\paragraph*{Size of \(\mathcal{S}\)}
Even if we restrict ourselves to small \(R\) values and thus short summaries, \(\mathcal{A}\) can still be large. The exponential relationship between the summary length and the size of \(\mathcal{S}\) makes it impractical to consider every possible summary and compute the distortion for all summary–text pairs. Therefore, we need a strategy to calculate the expected distortion between summaries and text without having to consider each pair.

\paragraph*{Finding \(d\)}
Defining the best similarity metric between a summary and text remains an active research topic. Proposed metrics range from ROUGE \cite{lin-2004-rouge} to more recent neural-based scores \cite{bert-score,bart-score}, LLM-based metrics \cite{song-etal-2024-finesure}, or information-theoretic scores \cite{darrin-etal-2024-cosmic}. They are typically judged by how well they correlate with human evaluations, which is considered the ground truth. However, since we require an automated procedure, we must adopt one of these existing metrics to compute the distortion.

To address the challenge of high-dimensional text space, we leverage LLMs to map the discrete text strings into a continuous embedding space. Because these models are trained on extensive corpora, they learn deep linguistic and semantic features represented in a Euclidean space, which often transfer well to downstream tasks \cite{bert}. By embedding texts into a continuous space, we can model the text source with a continuous distribution, enabling us to estimate \(p_T\) for unseen texts and compute \(\mathbb{E}[d(T,S)]\) easily without considering every summary-text pair.

Following the approach of reference \cite{kim2022mutual}, we fit a multivariate Gaussian distribution to the embeddings and adopt the squared Euclidean distance between the summary and text embeddings as our distortion function. This choice aligns with prior work \cite{bert-score,bart-score} which demonstrates that embedding-based similarity metrics correlate well with human judgments of textual similarity. In this setup, the analysis simplifies significantly thanks to the following theorem.

\begin{theorem} \label{thm:gaussian-sum-rd}
    Let $\mathbf{E}_T = (E_T^{(1)}, \dots, E_T^{(m)})$ be a random vector in $\mathbb{R}^m$ such that, given $\len(T) = l$, \(\mathbf{E}_T\) follows a Gaussian distribution whose covariance matrix has eigenvalues $\{\lambda_{l,i}\}_{i=1}^m$. Under the distortion measure
    \begin{equation}
        d(\mathbf{E}_T,\mathbf{E}_S) = \sum_{i=1}^m (E_T^{(i)} - E_S^{(i)})^2,
    \end{equation}
    the summarizer rate-distortion function is given by
    \begin{equation} \label{eq:ind-gauss-solution}
    R_S(D)=\frac{1}{\bar{L}}\sum_lp(l)\sum_{i=1}^m{\frac{1}{2}\log{\frac{\lambda_{l,i}}{D_{l,i}}}},
    \end{equation}
    where
    \begin{equation} \label{eq:ind-gauss-solution-condition}
        D_{l,i}=\begin{cases}
            c,& \text{if } c < \lambda_{l,i} \\
            \lambda_{l,i},& \text{if } c\geq \lambda_{l,i}
        \end{cases}
    \end{equation}
    and \(c\) is chosen such that \(\sum_l{p(l)\sum_{i=1}^mD_{l,i}}=D\).
\end{theorem}

\begin{IEEEproof}
    See Appendix~\ref{appendix:thm2}.
\end{IEEEproof}

In other words, under the Gaussian assumption, the summarizer rate-distortion function can be computed via reverse water-filling, which requires only the covariance matrix of the text embeddings for every $l$. To ensure sufficient data to calculate each sample covariance matrix, we define a length grid $\{l'_1, \dots, l'_n\}$ and merge all $\mathcal{T}_l$ for $l \in [l'_i, l'_{i+1})$ into $\mathcal{T}_{l'_i}$. This guarantees that each $\mathcal{T}_{l'_i}$ contains more than a specified number of texts. The full procedure for approximating $R_S(D)$ on a text dataset is outlined in Algorithm~\ref{alg:approx-rs-d}.

\begin{figure}[htbp]
    \centering
    \includegraphics[width=\linewidth]{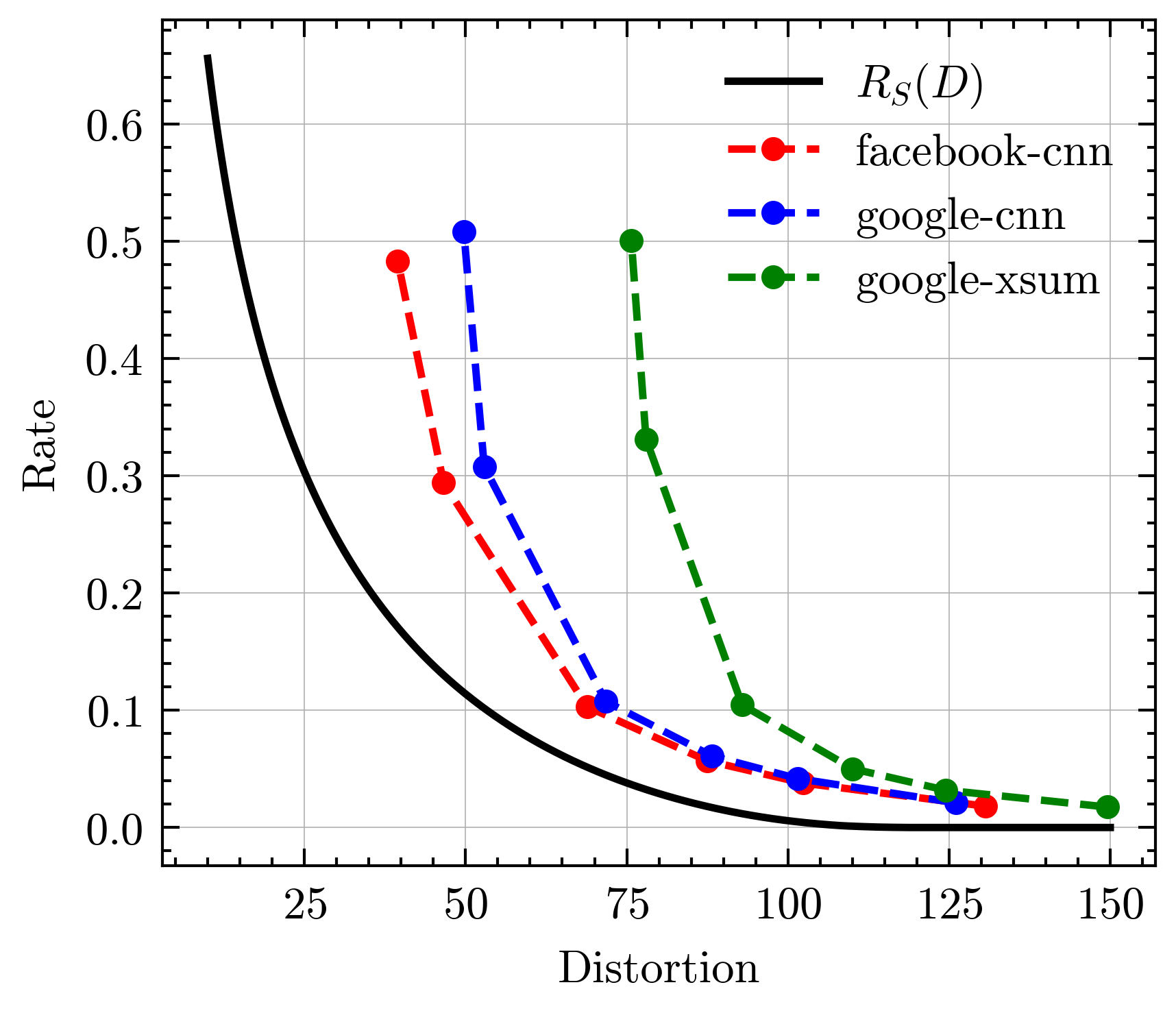}
    \caption{The summarizer rate-distortion curve, computed via Algorithm~\ref{alg:approx-rs-d} for CNN/DailyMail dataset, and the performance of different summarizers.}
    \label{fig:cnn-rd-curve}
\end{figure}

In Figure~\ref{fig:cnn-rd-curve}, we present the approximated \(R_S(D)\) for the CNN/DailyMail test split, computed from BGE-M3 embeddings \cite{bge_m3} under the Gaussian assumption. To compare this approximation with popular summarizers, we employ three widely used models from HuggingFace: \texttt{facebook/bart-\\large-cnn}, \;\texttt{google/pegasus-cnn\_dailymail} and \texttt{google/pegasus-xsum}. We run each summarizer at various rates on the same dataset and plot the resulting distortion-rate pairs in Figure~\ref{fig:cnn-rd-curve}. 

We observe that \texttt{facebook-cnn} and \texttt{google-cnn}, both fine-tuned on CNN/DailyMail, achieve a similar performance. By contrast, \texttt{google-xsum}, fine-tuned on a different dataset, is performing worse. This confirms that the squared distance between embeddings is a reliable distortion metric. We also see that the approximated summarizer rate-distortion curve serves as a plausible lower bound, but it becomes less accurate at low distortion levels. Notably, due to the continuous embedding space, it diverges to infinity at zero distortion. However, zero distortion can be achieved at \(R=1\) with the summarizer that simply returns the input text unchanged. More details about the experiment can be found in Appendix~\ref{appendix:experiment-details}.

\begin{algorithm}[ht]
\caption{Approximating $R_S(D)$ From a Dataset}
\label{alg:approx-rs-d}
\begin{algorithmic}[1]
\REQUIRE Dataset $\mathcal{D}$, Language Model $\mathcal{M}$, Distortion $D$
\vspace{2pt}
\STATE For each text $t \in \mathcal{D}$, generate embeddings $\mathbf{e}_{t} = \mathcal{M}(t)$
\STATE Create a length grid \(\{l'_1, l'_2, \dots, l'_n\}\)
\FOR{every \( i \in \{1,2,\dotsc,n-1\} \)}
    \STATE \( \mathcal{E}_i \gets \{ \mathbf{e}_{t} \mid \len(t) \in [l'_i, l'_{i+1}) \} \)
    \STATE \( \mathbf{K}_i \gets \text{SampleCovarianceMatrix}\left( \mathcal{E}_i \right) \)
\ENDFOR
\STATE \(R_S(D) \gets \text{ReverseWaterFilling}(\{\mathbf{K}_i\}_i, D)\)
\end{algorithmic}
\end{algorithm}

\section{Conclusion}
In this paper, we have proposed a rate-distortion framework for text summarization. Ongoing and future work includes incorporating perception measures to capture the naturalness of summaries and exploring how the lower bound can be helpful in evaluating different summarizers.

\IEEEtriggeratref{12}

\bibliographystyle{IEEEtran}
\bibliography{ref}

\clearpage
\appendices

\setcounter{theorem}{0}

\section{Appendix: Proof of Theorem \ref{thm:sum_bound1}}\label{appendix:thm1}
\begin{theorem}
    Let a summarizer \(p_{S^n|T^n}\) be such that \(\mathbb{E}[\len(S^n)|\len(T^n)] \leq R\len(T^n)\) with probability \(1\). If \(\mathbb{E}\left[d\left(T^n, S^n\right)\right] \leq D\), then \(R \geq R_S(D)\).
\end{theorem} 
\begin{IEEEproof}
Let \(\epsilon > 0\) and \(G=\mathds{1}\{\len(S^n) \leq R\len(T^n) + \epsilon n\}\). We first bound the conditional probability of the atypical event \(G=0\) given \(L^n = l^n\) using Chebyshev's inequality. Note that due to the additive structure of \(d(T^n, S^n)\) and \(\len(S^n)\), the optimal summarizer processes each \(T_i\) independently given \(L^n = l^n\). Therefore, without loss of generality, we consider conditionally independent \(\len(S_i)\), which yields

\begin{align}
    \Pr(G=0 | l^n)
        &= \Pr\left(\len(S^n) > R \sum_{i=1}^nl_i + \epsilon n \middle| l^n\right) \\
        &\leq \Pr\left(\len(S^n) > \mathbb{E}[\len(S^n)|l^n] + \epsilon n \middle| l^n\right) \\
        &\leq \frac{\mathrm{Var}[\len(S^n) | l^n]}{\epsilon^2n^2} \\
        &= \frac{\sum_{i=1}^{n}\mathrm{Var}[\len(S_i) | l^n]}{\epsilon^2n^2} \\
        &\leq \frac{1}{4\epsilon^2n^2}\sum_{i=1}^n{l_i^2} \label{eq:app_prob_bound},
\end{align}
where \eqref{eq:app_prob_bound} follows from the fact that \(\len(S_i) \leq \len(T_i)\) with probability \(1\).

Depending on \(G\), we can calculate the number of possible summaries as we know the alphabet and the upper bound on the length of the summary. This yields for the typical event

\begin{align}
    H(S^n|G=1, l^n) &\leq \log{\left|\text{Range}\left(S^n|G=1, l^n\right)\right|} \label{eq:entropy_bound_app} \\
    &\leq \log{\left(|\mathcal{A}|^{R \sum_{i=1}^nl_i + \epsilon n + 1}\right)-1}\\
    &\leq R \sum_{i=1}^nl_i + \epsilon n + 1 \label{eq:app_cond_ent1},
\end{align}
and similarly for the atypical event
\begin{equation}
    H(S^n|G=0, l^n) \leq \sum_{i=1}^nl_i + 1 \label{eq:app_cond_ent2}.
\end{equation}

Using \eqref{eq:app_prob_bound}, \eqref{eq:app_cond_ent1}, and \eqref{eq:app_cond_ent2}, we can bound the entropy of \(S^n\) conditioned on \(L^n=l^n\). Noting that \(H(G|S^n, l^n)=0\), we write
\begin{align}
    H(S^n|l^n) &= H(S^n|l^n) + H(G|S^n, l^n) \\
    &= H(S^n, G|l^n) \label{eq:app_chain_rule1} \\
    &= H(G|l^n) + H(S^n | G, l^n) \label{eq:app_chain_rule2} \\
    &\leq \log{2} + H(S^n | G, l^n) \label{eq:app_binary_ent} \\
    &= \log{2} + \Pr(G=1 \mid l^n) \, H(S^n \mid G=1, l^n) \notag \\
    &\quad + \Pr(G=0 \mid l^n) \, H(S^n \mid G=0, l^n) \\
    &\leq \log{2} + R\sum_{i=1}^nl_i + \epsilon n + 1 \notag \\
    &\quad + \left(\frac{1}{4\epsilon^2n^2}\sum_{i=1}^n{l_i^2}\right) \left(\sum_{i=1}^nl_i + 1\right),
\end{align}
where \eqref{eq:app_chain_rule1} and \eqref{eq:app_chain_rule2} follow from the chain rule, and \eqref{eq:app_binary_ent} follows from the fact that \(G\) is a binary random variable. Taking expectation over text lengths and using the finite third moment assumption yields
\begin{align}
    H(S^n|L^n) &= \sum_{l^n}p(l^n)H(S^n|l^n) \\
    &\leq \log{2} + Rn\bar{L} + \epsilon n + 1 \notag \\
    &\quad + \frac{(n^2-n)\mathbb{E}[\len(T)^2]\bar{L} + n\mathbb{E}[\len(T)^3]}{4\epsilon^2n^2} \notag \\
    &\quad+ \frac{n\mathbb{E}[\len(T)^2]}{4\epsilon^2n^2} \\
    &= n(\bar{L}R + \epsilon) + \mathcal{O}(1) \label{eq:ce_ub_app}, 
\end{align}
which can be used in a classical converse argument as follows:

\begin{align}
    n(\bar{L}R+\epsilon) + \mathcal{O}(1) &\geq H(S^n|L^n) \\
    &\geq I(T^n;S^n|L^n) \label{eq:non_negativity_app} \\
    &= H(T^n|L^n)-H(T^n|S^n,L^n) \\
    \begin{split}
        &= \sum_{i=1}^n{H(T_i|L^n, T_{i-1},\dotsc,T_1)} \\ &\qquad- \sum_{i=1}^n{H(T_i|S^n,L^n,T_{i-1},\dotsc,T_1)}
    \end{split} \label{eq:chain_rule_app} \\
    \begin{split}
        &= \sum_{i=1}^n{H(T_i|L_i)} \\ &\qquad- \sum_{i=1}^n{H(T_i|S^n,L^n,T_{i-1},\dotsc,T_1)}
    \end{split} \label{eq:text_independence_app} \\
    &\geq \sum_{i=1}^n{H(T_i|L_i)} - \sum_{i=1}^n{H(T_i|S_i,L_i)} \label{eq:cond_reduces_app} \\
    &= \sum_{i=1}^n{I(T_i;S_i|L_i)} \\
    &\geq \sum_{i=1}^n{\bar{L}R_S(\mathbb{E}[d(T_i, S_i)])} \label{eq:def_rd_app} \\
    &\geq n\bar{L}R_S(\sum_{i=1}^n{\frac{1}{n}\mathbb{E}[d(T_i, S_i)])} \label{eq:jensen_app} \\
    &= n\bar{L}R_S(\mathbb{E}[d(T^n, S^n)]) \\
    &\geq n\bar{L}R_S(D) \label{eq:final_ub_app},
\end{align}
where \eqref{eq:non_negativity_app} follows from the fact that \(H(S^n|T^n,L^n)\geq0\), \eqref{eq:chain_rule_app} follows from the chain rule, \eqref{eq:text_independence_app} follows from the independence of \(T_i\), \eqref{eq:cond_reduces_app} follows from the fact that conditioning reduces entropy, \eqref{eq:def_rd_app} follows from the definition of summarizer rate-distortion function, \eqref{eq:jensen_app} follows from Lemma~\ref{lemma:rd_convexity} and Jensen's inequality, and \eqref{eq:final_ub_app} follows from Lemma~\ref{lemma:rd_convexity}. The final bound implies \(R \geq R_S(D) - \frac{\epsilon}{\bar{L}} - \mathcal{O}\left(\frac{1}{n}\right)\).

Now suppose that \(R_S^{(n)}(D)\) is the lowest rate achieved summarizing \(T^n\). Observe that the subsequence \(\{R_S^{(k)}(D)\}_{k=n,2n,4n,\dotsc}\) must be non-increasing, since the summarizer for \(T^n\) can be applied to \(T^{2n}\) in two blocks to achieve the same rate and distortion. Therefore, \(R \geq R_S^{(n)}(D) \geq \liminf_{k\to\infty}R_S^{(k)}(D) \geq R_S(D) - \frac{\epsilon}{\bar{L}}\). Because \(\epsilon\) was arbitrary, we have \(R \geq R_S(D)\).
\end{IEEEproof}

\section{Appendix: Proof of Theorem \ref{thm:gaussian-sum-rd}}\label{appendix:thm2}
\begin{theorem}
    Let $\mathbf{E}_T = (E_T^{(1)}, \dots, E_T^{(m)})$ be a random vector in $\mathbb{R}^m$ such that, given $\len(T) = l$, \(\mathbf{E}_T\) follows a Gaussian distribution whose covariance matrix has eigenvalues $\{\lambda_{l,i}\}_{i=1}^m$. Under the distortion measure 
    \begin{equation}
        d(\mathbf{E}_T,\mathbf{E}_S) = \sum_{i=1}^m (E_T^{(i)} - E_S^{(i)})^2,
    \end{equation}
    the summarizer rate-distortion function is given by
    \begin{equation}
    R_S(D)=\frac{1}{\bar{L}}\sum_lp(l)\sum_{i=1}^m{\frac{1}{2}\log{\frac{\lambda_{l,i}}{D_{l,i}}}},
    \end{equation}
    where
    \begin{equation}
        D_{l,i}=\begin{cases}
            c,& \text{if } c < \lambda_{l,i} \\
            \lambda_{l,i},& \text{if } c\geq \lambda_{l,i}
        \end{cases}
    \end{equation}
    and \(c\) is chosen such that \(\sum_l{p(l)\sum_{i=1}^mD_{l,i}}=D\).
\end{theorem}

\begin{IEEEproof}
    First, consider the univariate case. Let \(E_T\) be a random variable such that given \(\len(T)=l\), it follows a Gaussian distribution with variance \(\sigma_l^2\). Then we have
    \begin{align}
        I(E_T;E_S|L=l)&=h(E_T|L=l)-h(E_T|E_S,L=l) \\
                      &=\frac{1}{2}\log{2\pi e \sigma_l^2} - h(E_T-E_S|E_S,L=l) \\
                      &\geq\frac{1}{2}\log{2\pi e \sigma_l^2} - h(E_T-E_S|L=l) \\
                      \begin{split}
                      &\geq \frac{1}{2}\log{2\pi e \sigma_l^2} \\ &\qquad- h(\mathcal{N}(0, \mathbb{E}[(E_T-E_S)^2|L=l])) \label{eq:gauss-bound}
                      \end{split} \\
                      &= \frac{1}{2}\log{2\pi e \sigma_l^2} - \frac{1}{2}\log{2\pi e D_l} \\
                      &=\frac{1}{2}\log{\frac{\sigma_l^2}{D_l}},
    \end{align}
    where \(D_l=\mathbb{E}[(E_T-E_S)^2|L=l]\) and \eqref{eq:gauss-bound} follows from the fact that the normal distribution maximizes the entropy for a given second moment. Since mutual information is always positive, we have
    \begin{equation}
        I(E_T;E_S|L=l)\geq \bigg(\frac{1}{2}\log{\frac{\sigma_l^2}{D_l}}\bigg)^+
    \end{equation}
    and
    \begin{equation}
        I(E_T;E_S|L)\geq \sum_lp(l)\bigg(\frac{1}{2}\log{\frac{\sigma_l^2}{D_l}}\bigg)^+.
    \end{equation}

    Now let $\mathbf{E}_T = (E_T^{(1)}, \dots, E_T^{(m)})$ be a random vector such that given $\len(T) = l$, each component $E_T^{(i)}$ is an independent Gaussian random variable with variance $\sigma_{l,i}^2$. Then we have
    \begin{align}
    I(\mathbf{E}_T;\mathbf{E}_S|L)&=h(\mathbf{E}_T|L)-h(\mathbf{E}_T|\mathbf{E}_S,L) \\
                    &\geq \sum_{i=1}^mh(E_T^{(i)}|L)-\sum_{i=1}^mh(E_T^{(i)}|E_S^{(i)},L) \label{eq:achieved-by-ind} \\
                    &=\sum_{i=1}^mI(E_T^{(i)};E_S^{(i)}|L) \\
                    &\geq \sum_{i=1}^{m}{\sum_lp(l)\bigg(\frac{1}{2}\log{\frac{\sigma_{l,i}^2}{D_{l,i}}}\bigg)^+} \label{eq:achieved-by-specific-cond} \\
                    &=\sum_lp(l)\sum_{i=1}^m{\bigg(\frac{1}{2}\log{\frac{\sigma_{l,i}^2}{D_{l,i}}}\bigg)^+},
    \end{align}   
    where \(D_{l,i}=\mathbb{E}[(E^{(i)}_T-E^{(i)}_S)^2|L=l]\). We can achieve equality in \eqref{eq:achieved-by-ind} by choosing \[p_{\mathbf{E}_T|\mathbf{E}_S,L=l}=\prod_{i=1}^mp_{E_T^{(i)}|E_S^{(i)},L=l}\] for every \(l\) and in \eqref{eq:achieved-by-specific-cond} by choosing the distribution as in Theorem~10.3.2 in \cite{elements_of_info_th}. Hence, \(R_S(D)\) can be found by solving the following optimization problem:
    \begin{equation}
    \begin{aligned}
    R_S(D)=\min_{\{D_{l,i}\}}\quad &\frac{1}{\bar{L}}\sum_lp(l)\sum_{i=1}^m{\bigg(\frac{1}{2}\log{\frac{\sigma_{l,i}^2}{D_{l,i}}}\bigg)^+}\\
    \textrm{s.t.} \quad & \sum_l{p(l)\sum_{i=1}^mD_{l,i}}\leq D.
    \end{aligned}
    \end{equation}
    By solving the KKT conditions, we obtain the following solution:
    \begin{equation} \label{eq:ind-gauss-solution-app}
        R_S(D)=\frac{1}{\bar{L}}\sum_lp(l)\sum_{i=1}^m{\frac{1}{2}\log{\frac{\sigma_{l,i}^2}{D_{l,i}}}},
    \end{equation}
    where
    \begin{equation} \label{eq:ind-gauss-solution-condition-app}
        D_{l,i}=\begin{cases}
            c,& \text{if } c < \sigma_{l,i}^2 \\
            \sigma_{l,i}^2,& \text{if } c\geq \sigma_{l,i}^2 
        \end{cases}
    \end{equation}
    and \(c\) is chosen such that \(\sum_l{p(l)\sum_{i=1}^mD_{l,i}}=D\).

    A Gaussian random vector can be transformed by a reversible linear transformation that diagonalizes its covariance matrix, resulting in independent components whose variances are the eigenvalues $\lambda_{l,i}$ of the covariance matrix. Therefore, for a general Gaussian, the solution remains valid if we replace $\sigma_{l,i}^2$ with $\lambda_{l,i}$ in \eqref{eq:ind-gauss-solution-app} and \eqref{eq:ind-gauss-solution-condition-app}.
\end{IEEEproof}
\pagebreak
\section{Appendix: Experiment Details}\label{appendix:experiment-details}

\subsection{Tokenization}
Tokenization, in the context of natural language processing, refers to the process of splitting the input text into smaller units called tokens (e.g., words, subwords or characters). A popular approach to identify the unique tokens is byte-pair encoding, which merges frequent subword pairs iteratively. As a rule of thumb, around 0.75 words correspond to one token in English. The set of unique tokens define the alphabet of the tokenizer and the language model that employs this tokenizer. Typically, every language model has its own tokenizer and a different tokenization procedure.

\subsection{The Embedding Model}
Any dataset used for summarization is expected to contain long texts. If we want to represent texts of varying lengths in the same embedding space, the language model has to be able to handle long inputs. Moreover, the output embedding dimension should be large enough to capture the intricacies of the large text space. Because of its remarkable ability to handle long texts, we choose to use BGE-M3 model, which can be accessed on HuggingFace with the path \texttt{BAAI/bge-m3}. 

\begin{figure}[htbp]
    \centering
    \includegraphics[width=\linewidth]{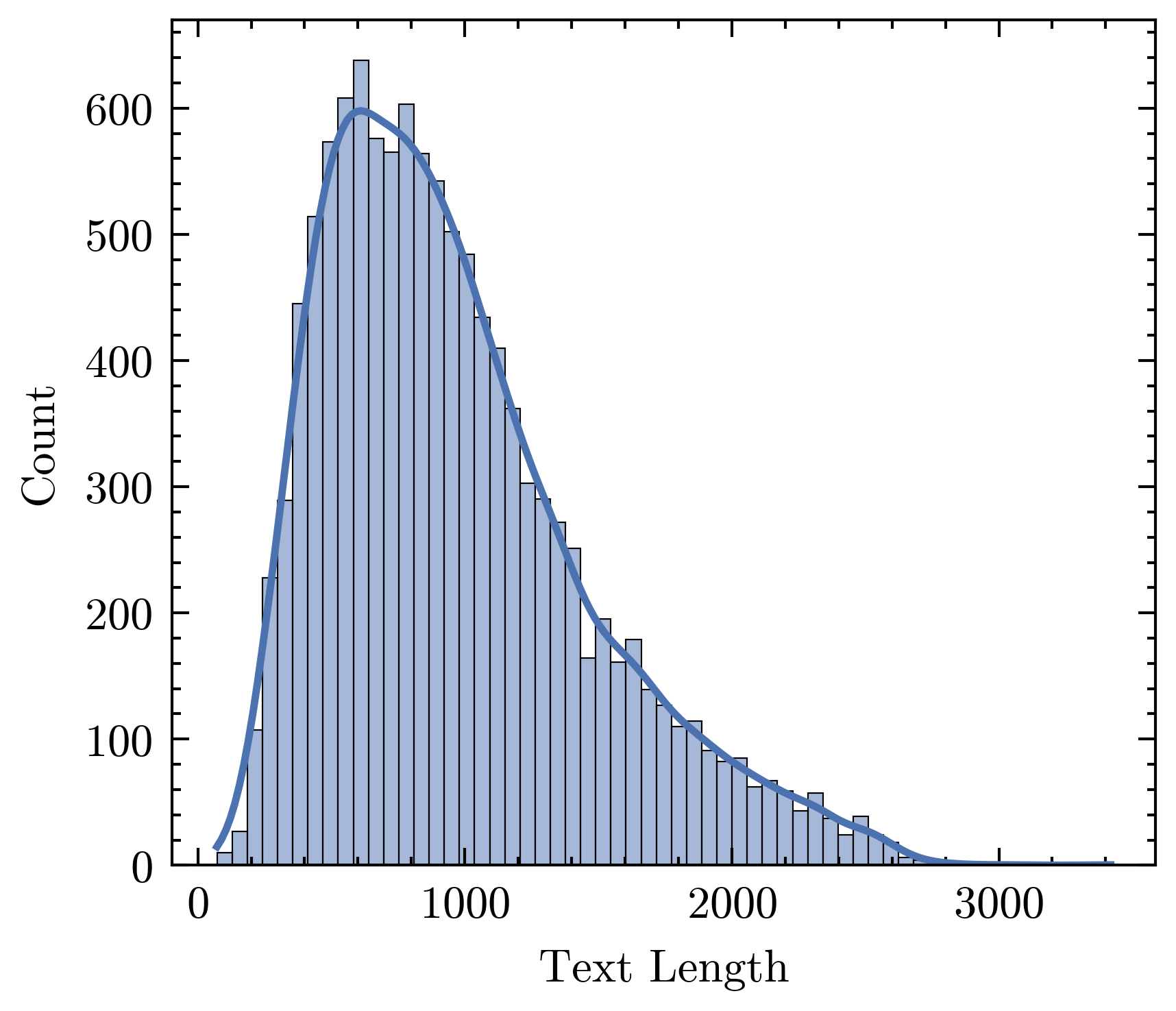}
    \caption{The distribution of text lengths in the CNN/DailyMail test split, tokenized with BGE-M3 tokenizer}
    \label{fig:cnn-length-hist}
\end{figure}

BGE-M3 model can process input sequences up to \(8192\) tokens. With its tokenizer, the average length of the CNN/DailyMail dataset becomes \(965\) tokens, which can be handled easily by the model. The distribution of text lengths in the CNN/DailyMail test split is shown in Figure~\ref{fig:cnn-length-hist}. Furthermore, the embedding dimension of BGE-M3 model is \(1024\), which is large enough to capture the nuanced semantic relationships present in lengthy texts.

BGE-M3 tokenizer contains \(250\text{k}\) unique tokens but most of these tokens are coming from other languages as it is a multi-lingual model. If we only consider the unique tokens that occur in the CNN/DailyMail dataset this number drops to \(30\text{k}\), which becomes our alphabet size.

\subsection{The Pooling Method}
The output of an embedding model is the contextual embeddings for each token in the input sequence. To obtain a single embedding vector for the entire text, a pooling method needs to be applied. Among the popular pooling methods are CLS pooling and mean pooling. CLS pooling refers to using the embedding of the special \texttt{[CLS]} token, which is added to the beginning of every sequence and presumed to capture the semantics of the entire input text. Mean pooling refers to calculating the average of all the token embeddings.

We found that using the mean pooling method results in text embeddings that have very low variance. We argue that this is because many tokens repeat across different texts and taking the average of token embeddings reduces the overall variance due to the common tokens in each text. Consequently, the eigenvalues of the resulting covariance matrices are underestimated, producing a lower summarizer rate-distrtion curve. To address this issue, we employ CLS pooling method instead.

Usually, a normalization step follows the pooling procedure, such as scaling the text embeddings to have an L2 norm of 1. This is helpful if the embeddings will be used to measure similarities (e.g. cosine similarity score). However, since we use squared distance as the distortion function, the normalization has a crucial effect on the achieved distortion values. In addition, normalization reduces the variance of the embeddings as well. Hence, we do not apply normalization after pooling.

\subsection{Creating The Length Grid}
Since we want to calculate the sample covariance matrix of vectors of dimension 1024, we choose to use at least 2k samples to have a stable estimate. The CNN/DailyMail test split contains 11,490 news articles and it can be seen from Figure~\ref{fig:cnn-length-hist} that length distribution is right skewed. Hence, more texts must be merged at longer lengths to have equal number of samples. Our resulting length grid is \(\{71,\,588,\,866,\,1233,\,3419\}\), which yields about 2,800 texts per interval.

\subsection{Eigenvalues of Covariance Matrices}
\begin{figure*}[htbp]
    \centering
    \includegraphics[width=0.8\linewidth]{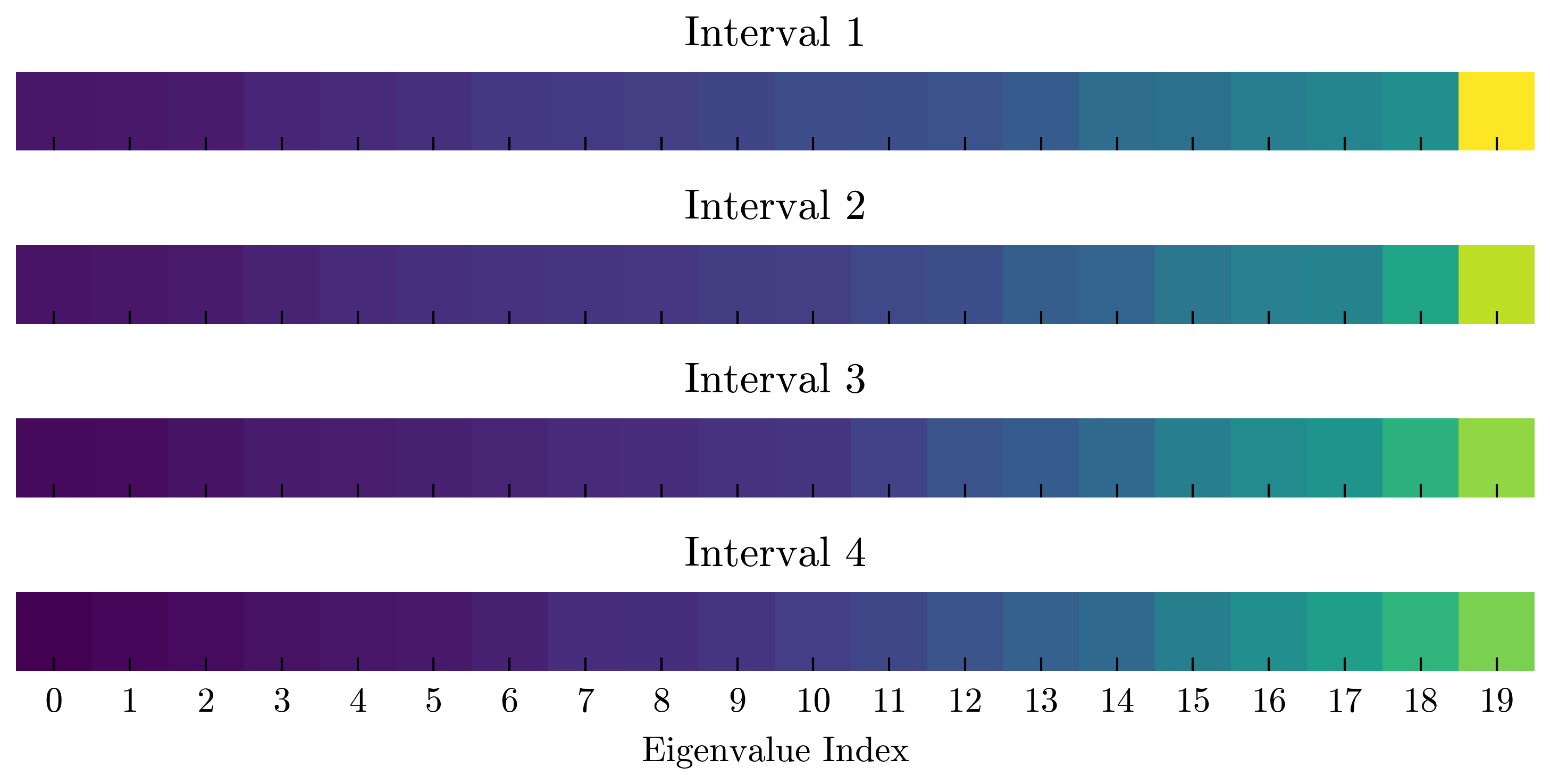}
    \caption{Comparison of the sample covariance eigenvalues for each interval.}
    \label{fig:cov-eigvalues}
\end{figure*}

\begin{figure}[htbp]
    \centering
    \includegraphics[width=\linewidth]{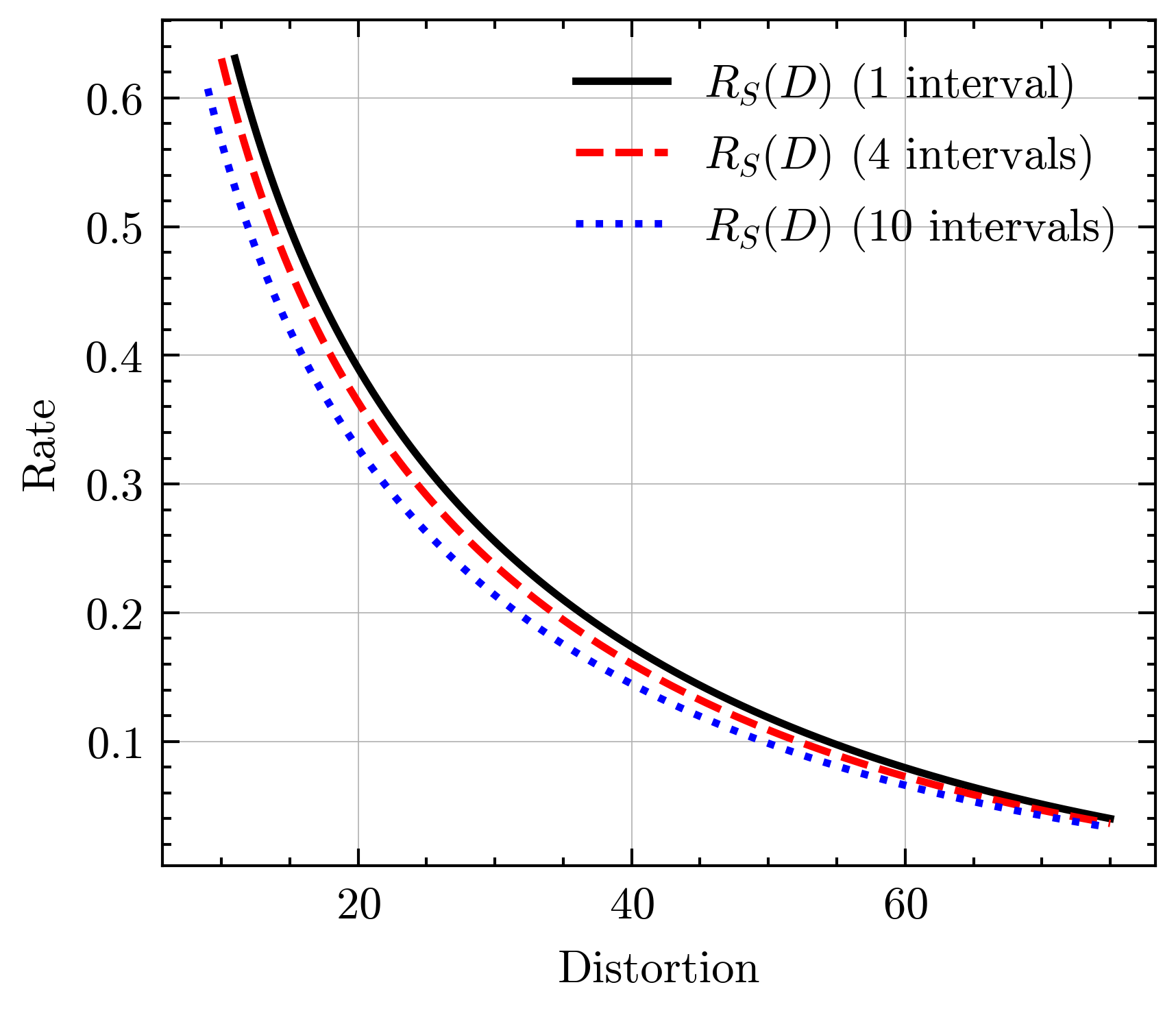}
    \caption{Comparison of summarizer rate-distortion curves computed with Algorithm~\ref{alg:approx-rs-d} using different number of length intervals.}
    \label{fig:sum-rd-different-intervals}
\end{figure}

We compare the top 20 sample covariance eigenvalues of the text embeddings in each interval in Figure~\ref{fig:cov-eigvalues}. We observe that the largest eigenvalues are nearly identical, suggesting \(\lambda_{l,i} \approx \lambda_i\) for every \(l\). Substituting \(\lambda_{l,i}=\lambda_i\) in \eqref{eq:ind-gauss-solution} and \eqref{eq:ind-gauss-solution-condition} reduces Theorem~\ref{thm:gaussian-sum-rd} to the classical reverse water filling for a multivariate Gaussian with eigenvalues \(\lambda_i\). Consequently, calculating the summarizer rate-distortion curve can be further simplified by making this assumption. This also corresponds to defining the length grid to be only the minimum and the maximum text length, i.e., single interval, if the principal components for every \(l\) are in the same direction. In Figure~\ref{fig:sum-rd-different-intervals}, we compare the summarizer rate-distortion curves calculated using 10 intervals, 4 intervals, and a single interval.

\subsection{Running Summarizers At Specific Rates}
Summarizer models require specifying minimum and maximum summary lengths, but these lengths are calculated based on each summarizer's own tokenizer. By contrast, the rate is computed as in Definition~\ref{def:summarizer} with the embedding model's tokenizer to ensure a fair comparison. Hence, although the intended rates were set to \(\{0.5,\, 0.3,\, 0.1,\, 0.05,\, 0.03,\, 0.01\}\), it can be seen from Figure~\ref{fig:cnn-rd-curve} that the actual rates deviate slightly from these values.

\subsection{The Efficacy of Algorithm~\ref{alg:approx-rs-d}}
Finally, we would like to point out that the accuracy of the summarizer rate-distortion function computed with Algorithm~\ref{alg:approx-rs-d} depends on the answers to the following questions:
\begin{enumerate}
    \item How accurately can the chosen language model capture the dataset’s 
          semantics in the embedding space?
    \item To what extent do the embeddings follow a Gaussian distribution?
    \item How reliably can the sample covariance matrices be estimated from 
          the dataset?
\end{enumerate}

\end{document}